\documentclass[aip,jap,amsmath,amssymb,
reprint,
footinbib,unsortedaddress]{revtex4-1}
\usepackage{CJK}
\usepackage{dcolumn}
\usepackage{graphicx}
\usepackage{amssymb}
\usepackage{bm}
\usepackage{dcolumn}
\usepackage{bm}
\usepackage[pdftex]{hyperref}

\begin{document}
\begin{CJK*}{SJIS}{MS Mincho} 

\title{Fabrication of submicron La$_{2-x}$Sr$_{x}$CuO$_{4}$ intrinsic Josephson junction stacks}

\author{Yuimaru Kubo$^{1,2}$}\thanks{Present address: Service de Physique de l'Etat Condens\'e (SPEC), CEA-Saclay, Gif-sur-Yvette 91191, France.}

\author{Yamaguchi Takahide$^{1}$}

\author{Takayoshi Tanaka$^{3}$}

\author{Shinya Ueda$^{1}$}\thanks{Present address: Department of Applied Physics, Tokyo University of Agriculture and Technology, Tokyo 184-8588, Japan.}

\author{Satoshi Ishii$^{1}$} \thanks{Present address: Department of Quantum Engineering, Nagoya University, Nagoya 464-8603, Japan}

\author{Shunsuke Tsuda$^{1}$}

\author{ATM Nazmul Islam$^{4}$}\thanks{Present address: Helmholtz-Zentrum Berlin f\"ur Materialien und Energie GmbH, 14109 Berlin, Germany}

\author{Isao Tanaka$^{4}$}

\author{Yoshihiko Takano$^{1,2}$}

\affiliation{$^{1}$Superconducting Materials Center, National Institute for Materials Science (NIMS), Tsukuba 305-0047, Japan}
\affiliation{$^{2}$Graduate School of Pure and Applied Sciences, University of Tsukuba, Tsukuba 305-8571, Japan}
\affiliation{$^{3}$Material Manufacture and Engineering Station, National Institute for Materials Science (NIMS), Tsukuba 305-0047, Japan}
\affiliation{$^{4}$Center for Crystal Science and Technology, University of Yamanashi, Kofu 400-8511,Yamanashi, Japan}

\date{\today}
\begin{abstract}
Intrinsic Josephson junction (IJJ) stacks of cuprate superconductors have potential to be implemented as intrinsic phase qubits working at relatively high temperatures. 
We report success in fabricating submicron La$_{2-x}$Sr$_{x}$CuO$_{4}$ (LSCO) IJJ stacks carved out of single crystals. 
We also show a new fabrication method in which argon ion etching is performed after focused ion beam etching. 
As a result, we obtained an LSCO IJJ stack in which resistive multi-branches appeared. 
It may be possible to control the number of stacked IJJs with an accuracy of a single IJJ by developing this method.

\end{abstract}
\maketitle

\end{CJK*}

\section{Introduction}
Intrinsic Josephson junctions (IJJs) have been regarded as high quality Josephson junctions because of their atomically flat structure\cite{Kleiner:PRB1994}. 
Macroscopic quantum tunneling (MQT) has been reported in small IJJ stacks in recent years \cite{Inomata:PRL2005, Jin:PRL2006, SXLi:PRL2007, Kashiwaya:JPSJ2008, Ota:PRB2009, Ueda:JAP2009, Kubo:APEX2010} and has attracted attention\cite{Kawabata:PRB2004, Yokoyama:PRB2007, Machida:SST2007, Koyama:PC2008, Savelev:PRB2008} because of the existence of new quantum phenomena\cite{Jin:PRL2006} and the possibility of an ``intrinsic phase qubit'' working at higher temperatures. 

Bi$_{2}$Sr$_{2}$CaCu$_{2}$O$_{8+y}$ (BSCCO) has often been chosen for those MQT studies, because one can easily cleave the crystals with Scotch tape to prepare thin and flat crystals. 
In fact obtaining thin and flat crystals is the crucial point in the fabrication process to form a small $s$-shaped IJJ stack either using focused ion beam (FIB) etching technique\cite{Kim:SST1999} or using double sided cleaving method, i.e., argon ion etching\cite{Wang:APL2001,You:APL2006}. 
Because of this, much less experimental studies using IJJ stacks of other cuprates have been done. 

La$_{2-x}$Sr$_{x}$CuO$_{4}$ (LSCO) is known as a material having large Josephson plasma frequency $\omega_{p}$ reaching THz\cite{Tamasaku:PRL1992, Uchida:PRB1996}. 
The quantum fluctuations in phase space are proportional to $\omega_{p}$ \cite{Martinis:PRB1987, Wallraff:RSI2003}; a Josephson junction having larger $\omega_{p}$ may let us perform qubit operations at higher temperature. 
Therefore LSCO is one of the favorable materials to be implemented as intrinsic phase qubits. 
However an LSCO IJJ stack cannot easily be fabricated since it is difficult to cleave the bulk crystal by Scotch tape, unlike BSCCO. 
This means that the double sided cleaving technique\cite{Wang:APL2001, You:APL2006} cannot be applied to LSCO. 
Indeed only a few experimental studies fabricating $s$-shaped LSCO IJJ stacks in bulk crystals have been reported \cite{Uematsu:PC2001, Kim:IEEE2005, Kubo:PC2008, Kitano:PC2009}. 
Another possibility is to use thin films but \textit{a priori} this is not a good solution because the grain boundaries cannot be neglected\cite{Mizugaki:JAP2003, Ota:Thesis}.

In this article, (1) we report success in fabricating an LSCO IJJ stack in a bulk crystal with \textit{submicron lateral dimensions} by FIB etching, and (2) we propose a new method to fabricate a submicron LSCO IJJ stack using both \textit{FIB etching and argon ion etching}. 
The IJJ stacks have \textit{the smallest lateral dimensions} reported so far, and the new fabrication method allows us to \textit{control the number} $N_{\textrm{IJJ}}$ \textit{of stacked LSCO IJJs} with high precision. 
This new method could also be applied to other cuprate materials such as REBa$_{2}$Cu$_{3}$O$_{7-y}$ (RE: rare earth) compounds, which can hardly be cleaved by Scotch tape, unlike BSCCO.


\section{Sample Preparation}\label{sec:Cut}

\begin{figure}
\begin{center}
\includegraphics[width=0.8\hsize]{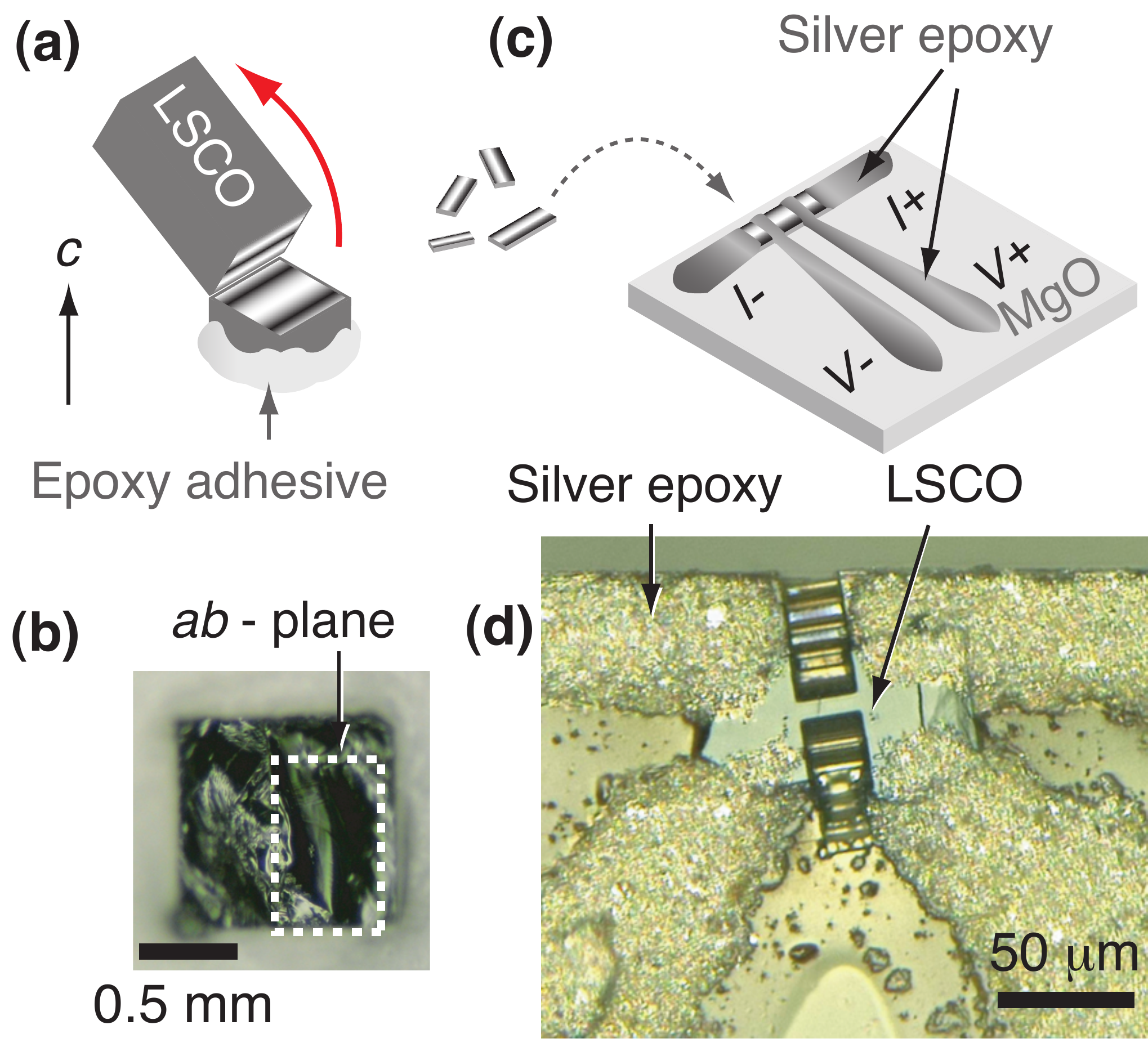}
\caption{(Color online) Schematic of sample preparation.  (a) An LSCO rod is cleaved. The long direction is aligned to the $c$-axis.
(b) An optical microscope image after cleaving.
A shiny surface of the $ab$-plane is obtained as shown inside the dotted square. 
(c) Tiny specimens are cut out from the cleaved rod by a wire saw, and one of the tiny pieces is fixed onto the substrate by polyimide adhesive. 
Silver epoxy is glued on the LSCO specimen to make four electrodes. (d) An optical microscope image of a sample. This picture was taken after a micro bridge was formed [see Fig. \ref{FIBetching.pdf}(a)].}
\label{Cleaving.pdf} 
\end{center}
\end{figure}

To carve out an $s$-shaped IJJ stack in an LSCO single crystal using the FIB etching technique\cite{Kim:SST1999}, we need to prepare a specimen which is thin along the $c$-axis.  
The desired thickness is $\lesssim$ 30 $\mu$m, because of the slow etch rate of the FIB system\cite{FIB}. 
Figure \ref{Cleaving.pdf} shows the process to prepare thin LSCO specimens.  
An LSCO single crystal grown by the traveling solvent floating zone method \cite{Tanaka:N1989}, with nominal Sr concentration $x =$  0.09, was cut and formed into a small rectangular parallelepiped shape. 
The dimensions are about 1 $\times$ 1 $\times$ 3 mm$^{3}$, and the long edge is along the $c$-axis [Fig. \ref{Cleaving.pdf}(a)]. 
The direction of the crystal $c$-axis was determined by X-ray Laue diffraction patterns before cutting. 

The small rod was fixed onto a metal plate with epoxy adhesive (Torr Seal, Ideal Vacuum Products), and carefully cleaved as schematically depicted in Fig. \ref{Cleaving.pdf}(a).  
A shiny flat surface, an indication of the $ab$-plane, appeared after cleaving as shown in Fig. \ref{Cleaving.pdf}(b).  
Tiny crystal pieces were obtained by cutting the flat part using a wire saw.  
The typical lateral dimension of these pieces is 0.05 - 0.1 mm, and the typical thickness is 20 - 50 $\mu$m.  
One of the pieces was fixed on a magnesium oxide (MgO) substrate by polyimide glue, and four electrical contacts were made by silver epoxy as shown in Fig. \ref{Cleaving.pdf}(c) and \ref{Cleaving.pdf}(d).  
To reduce contact resistance, we annealed the sample at 400 $^{\circ}$C for 10 minutes in oxygen in a mirror furnace (MILA-3000, Ulvac).

\section{Fabrication of a Sub-micron LSCO IJJ Stack by FIB Etching}\label{sec:FIB}

\begin{figure}
\begin{center}
\includegraphics[width=0.9\hsize]{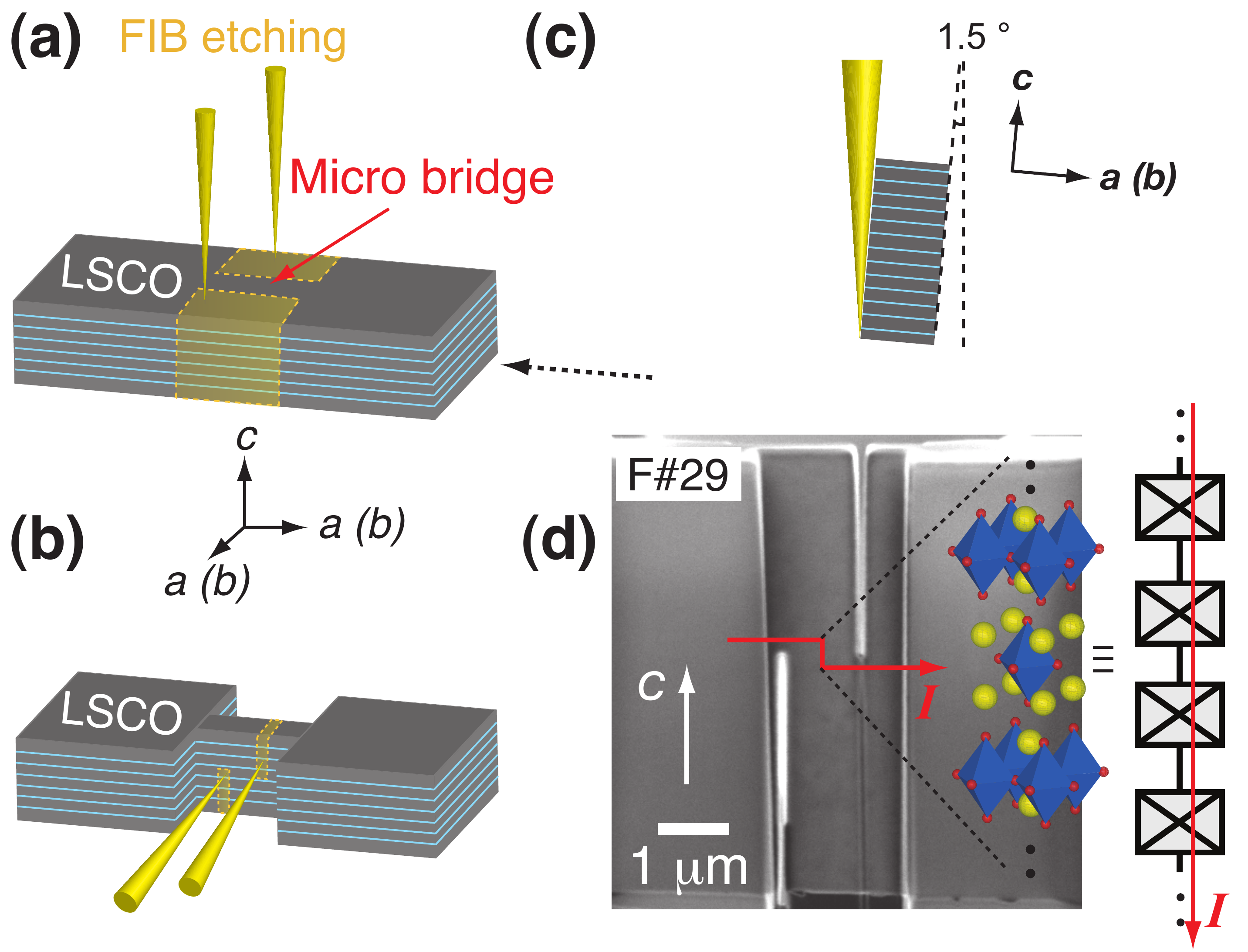}
\caption{(Color online) Schematic of FIB process. 
(a) Fabrication of a micro bridge by FIB etching.
(b) Two slits are formed by FIB etching from the $a (b)$-axes direction. 
(c) A technique to fabricate a fine structure using FIB etching.  
This picture shows a cross-section of the micro bridge viewed from the direction of the dotted arrow in (a). 
The micro-bridge is slightly ($\sim 1.5^{\circ}$) tilted to compensate for the finite angle of the ion beam. 
(d) A scanning ion microscope image of the IJJ stack ``F\#29''.  
The right hand shows the crystal structure of LSCO and a schematic of the IJJ array. 
}
\label{FIBetching.pdf}
\end{center}
\end{figure}

We used a focused a FIB/scanning ion microscope (FIB/SIM) system (SMI-9800SE, Seiko instruments) for both etching and observation. 
First a constricted micro-bridge was formed in the center of the sample as shown in Fig. \ref{FIBetching.pdf}(a). 
The length of the bridge is about 5 $\mu$m and the width is about 1 $\mu$m. 
Then we etched the micro-bridge from the $ab$ direction in order to make two slits, as depicted in Fig. \ref{FIBetching.pdf}(b). 
The two slits are separated by $\sim$ 0.9 $\mu$m. 
Finally the micro-bridge was etched from the $c$ direction again, to narrow the width down to $\sim$ 0.5 $\mu$m. 
Here the beam angle is slightly ($\sim$ 1.5$^{\circ}$) tilted to compensate for the finite angle of the ion beam, as schematically shown in Fig. \ref{FIBetching.pdf}(c). 

Since the two slits overlap slightly along the $c$-axis, an applied current $I$ flows along the $c$ direction in-between the slits as shown by the red arrow in Fig. \ref{FIBetching.pdf}(d). 
Therefore the small area can be regarded as an $s$-shape IJJ stack \cite{Kim:SST1999,Wang:APL2001,You:APL2006} as schematically depicted on the right hand side in Fig. \ref{FIBetching.pdf}(d). 
The lateral dimension of the LSCO IJJ stack ``F\#29'' is 0.45 $\times$ 0.95 $\mu$m$^{2}$.
The thickness is about 30 - 40 nm, estimated from SIM images. 
To our knowledge, this is the first report of a \textit{submicron LSCO IJJ stack} formed in a bulk crystal.

\section{Discussions for F\#29}


\begin{figure}
\begin{center}
\includegraphics[width=0.7\hsize]{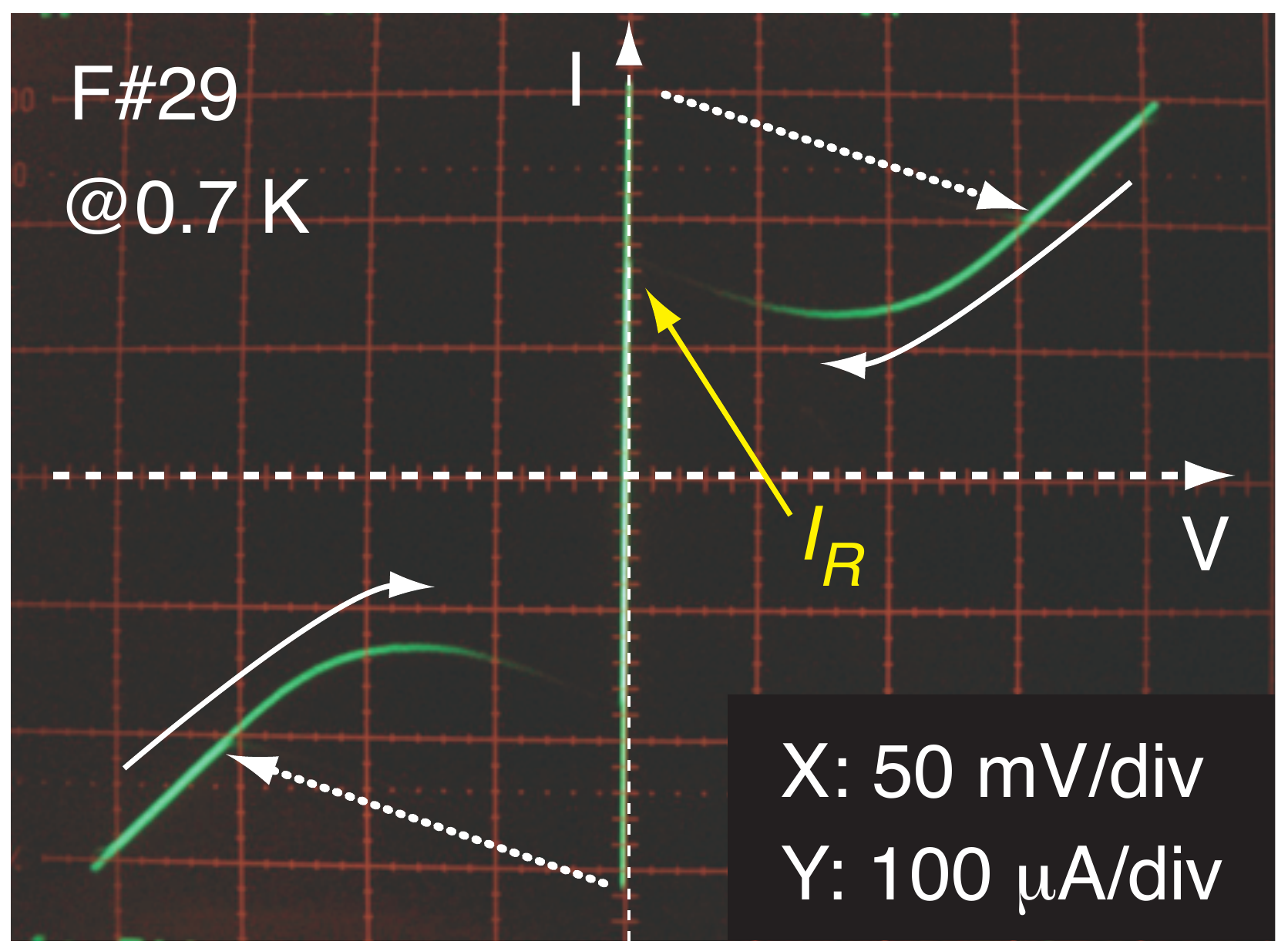}
\caption{(Color online) A snapshot of an $I - V$ curve of the LSCO IJJ stack \#F29.
The arrows indicate the course of the current ramp. 
The retrapping current $I_{R}$ is represented by the yellow arrow.} 
\label{IVLSCO29.pdf}
\end{center}
\end{figure}

Figure \ref{IVLSCO29.pdf} shows a current-voltage ($I - V$) characteristic of F\#29 at 0.7 K. 
A sharp voltage jump (the dashed arrows in Fig. \ref{IVLSCO29.pdf}) and a large hysteresis (the solid arrows in Fig. \ref{IVLSCO29.pdf}) are observed. 
These features suggest that the small structure fabricated acts as a stack of tunnel Josephson junctions, similar to BSCCO IJJ stacks\cite{Kleiner:PRB1994, Wang:APL2001}. 
Therefore F\#29 can be regarded as a submicron LSCO IJJ stack.

Resistive multi-branches\cite{Kleiner:PRB1994} do not appear in F\#29. 
On the contrary, we recognize that all the junctions simultaneously switch to their resistive state (dotted arrows in Fig. \ref{IVLSCO29.pdf}). 
This fact implies that all the IJJs have the same critical currents; therefore our LSCO crystal has homogeneous carrier density, and the LSCO IJJ stack was fabricated without any extra damage such as residual gallium ions. 
A similar switching phenomenon has been already reported in uniform BSCCO stacks, and referred to as ``uniformly switching'' by Jin \textit{et al.} \cite{Jin:PRL2006}. 
As the great enhancement of the MQT rate proportional to $N_{\textrm{IJJ}}^{2}$ ($N_{\textrm{IJJ}}$: the number of stacked IJJs) was observed in the BSCCO uniform stacks\cite{Jin:PRL2006}, the unconventional and interesting quantum phenomenon may also emerge in LSCO IJJ stacks. 

Another feature of F\#29 is the large retrapping current $I_{R}$ (see the arrow in Fig. \ref{IVLSCO29.pdf}). 
In underdamped JJ's, the fluctuation free retrapping current $I_{R0}$ is given by \cite{Krasnov:PRB2007} 
\begin{equation}\label{Eq:Retrapping}
I_{R0} \simeq \frac{4I_{c0}}{\pi Q}, 
\end{equation}
where $I_{c0}$ is the fluctuation free critical current, and $Q$ is the quality factor of the junction. 
From Eq. \ref{Eq:Retrapping} and Fig. \ref{IVLSCO29.pdf}, $Q$ of F\#29 is roughly estimated to be 3 - 4. 
This means that the LSCO IJJ stack is moderately damped. 
Discussing about the damping effect in the LSCO IJJ stack here is beyond the scope of this article. 
The damping effects of BSCCO IJJ stacks as well as various types of Josephson junctions have been intensively studied by Krasnov \textit{et al} \cite{Krasnov:PRB2007}.

\section{Fabrication of an LSCO IJJ Stack By FIB and Argon Ion etching}


\begin{figure}
\begin{centering}
\includegraphics[width=0.9\hsize]{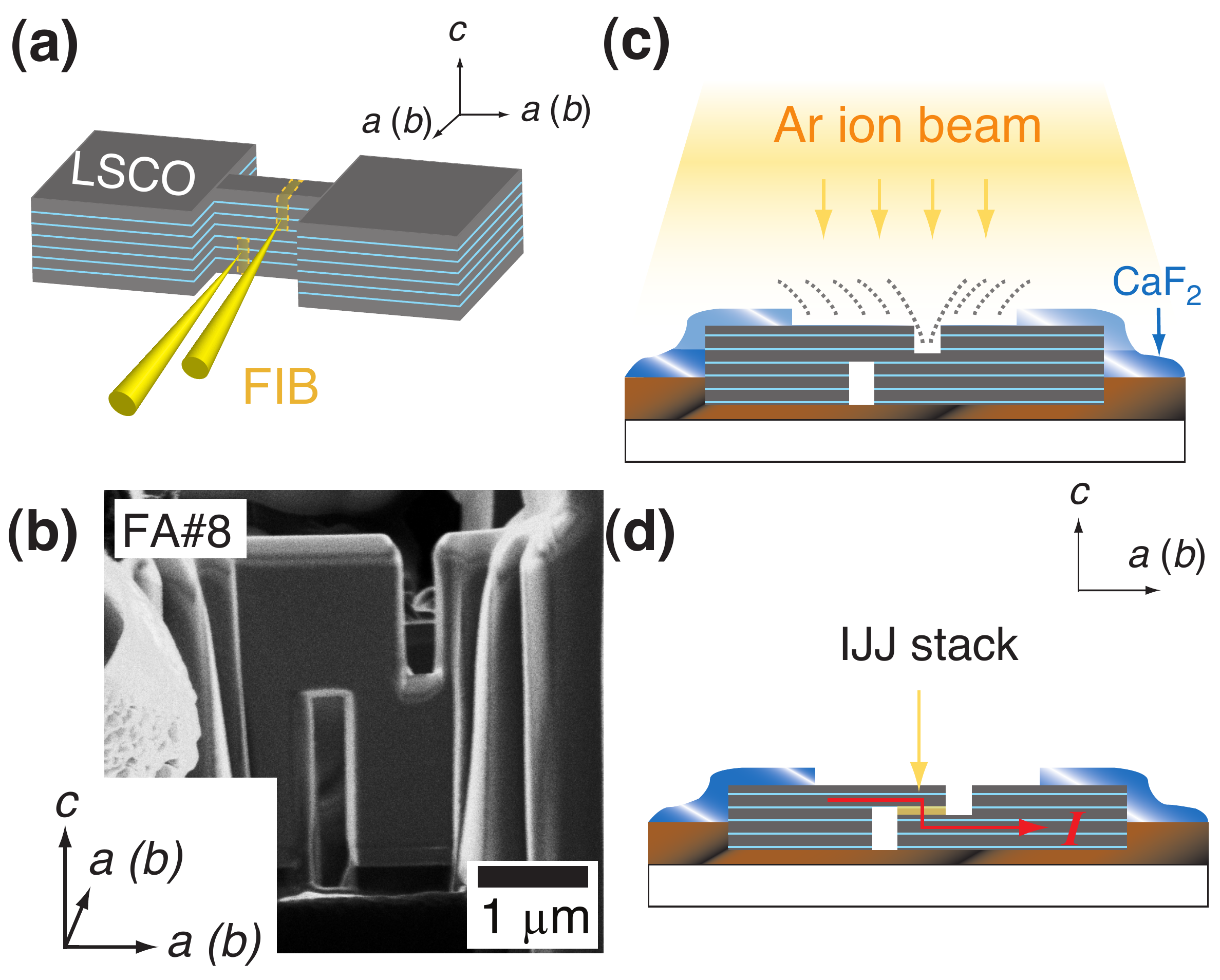}
\caption{(Color online) A schematic of the developed process.  
(a) FIB etching was performed from the direction of the $a (b)$-axes to form two slits.  
Here the two slits do not overlap along the $c$-axis. 
(b) A SIM image of ``FA\#8'' [taken after the process (a)]. 
(c) Argon ion etching to develop the upper slit.
(d) An LSCO IJJ stack with a small number of junctions is formed.  
}
\label{FIBandmilling.pdf}
\end{centering}
\end{figure}

In section \ref{sec:FIB}, we described the fabrication process for an LSCO IJJ stack using only FIB etching. 
A typical spatial resolution of the FIB machine, however, is about 10 - 50 nm. 
This is rather coarse with respect to the thickness of a single IJJ of LSCO, $\approx$ 0.7 nm. 
Therefore it is extremely difficult to control $N_{\textrm{IJJ}}$ using the FIB machine. 
Because of the new quantum phenomenon mentioned earlier\cite{Jin:PRL2006}, controlling $N_{\textrm{IJJ}}$ of an LSCO IJJ stack is an interesting although difficult challenge. 
Here, we present a developed fabrication method in which  argon ion etching is performed after FIB etching in order to gradually increase the thickness of the LSCO IJJ stack.

After making a micro bridge with 0.7 $\mu$m width by the same procedure described in section \ref{sec:FIB} and Fig. \ref{FIBetching.pdf}(a), calcium fluoride (CaF$_{2}$) was evaporated through a metal mask in order to make a protection layer for the electrodes. 
Then two slits were formed by FIB etching as schematically depicted in Fig. \ref{FIBandmilling.pdf}(a). 
In Fig. \ref{FIBandmilling.pdf}(b) we show a SIM image of the sample ``FA\#8'' taken after the process of Fig. \ref{FIBandmilling.pdf}(a). 
The distance between the slits is 0.5 $\mu$m. 
Note that the two slits do not overlap along the $c$-axis, i.e., the IJJ stack is not yet formed at this point. 
In the next step, the depth of the top slit was gradually increased using an argon ion etching machine (MPS-3000, Ion tech) as schematically shown in Fig. \ref{FIBandmilling.pdf}(c). 
As a result, a submicron LSCO IJJ stack with small $N_{\textrm{IJJ}}$ is formed as depicted in the center of Fig. \ref{FIBandmilling.pdf}(d). 

\section{Discussions for FA\#8}\label{sec:discF8}


\begin{figure}
\begin{centering}
\includegraphics[width=0.7\hsize]{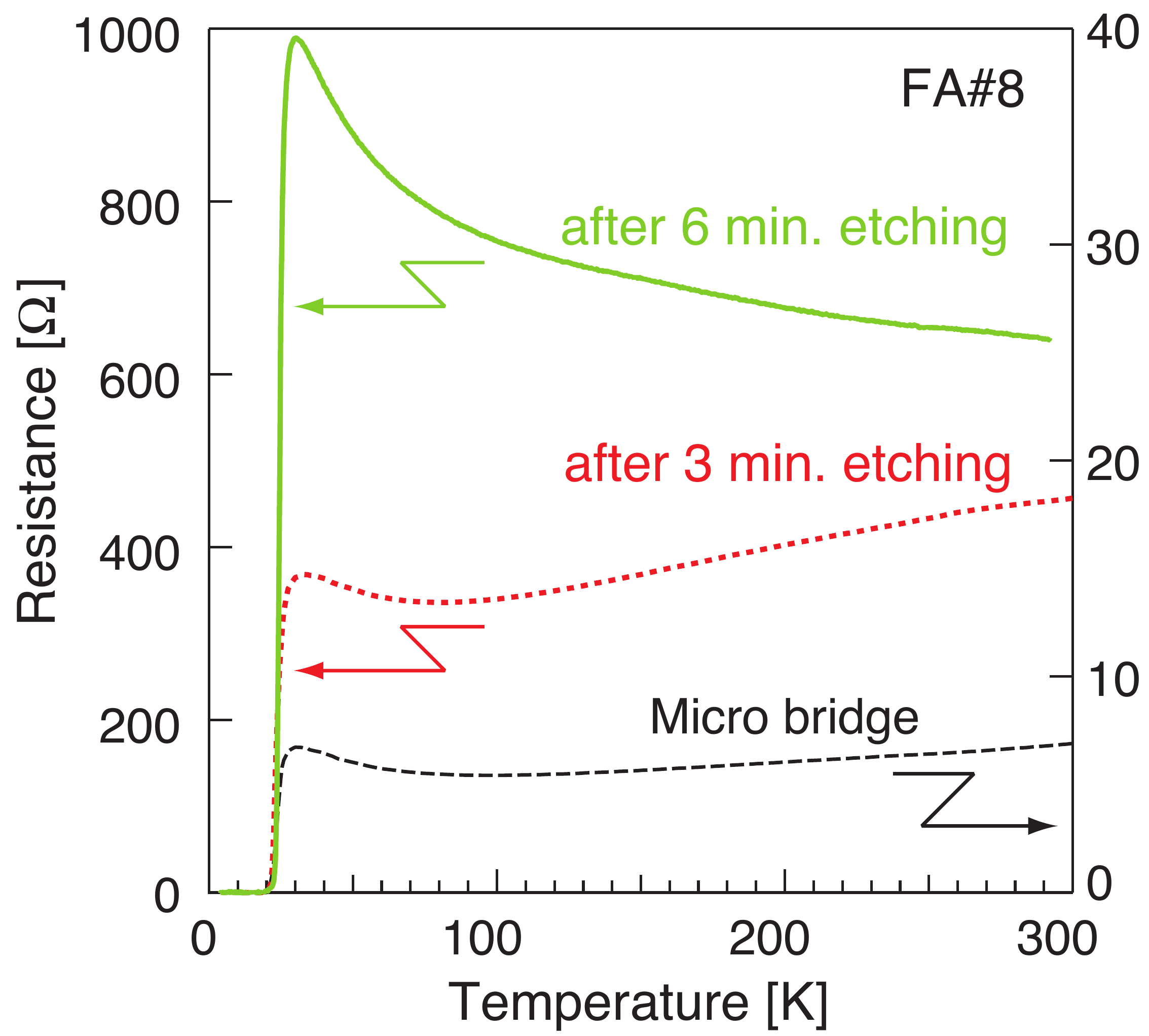} 
\caption{(Color online) 
Evolution of the resistance of FA\#8. 
The black dashed curve is of the micro-bridge, as of Fig. \ref{FIBetching.pdf}(a). 
The red dot curve is of after the first argon etching for 3 min. 
The green solid curve is of after the second argon etching for 3 more min. 
The right axis is the scale for the black dashed curve while the left is for the red dotted and the green solid curves. 
}
\label{RT.pdf} 
\end{centering}
\end{figure}

\begin{figure}
\begin{centering}
\includegraphics[width=0.8\hsize]{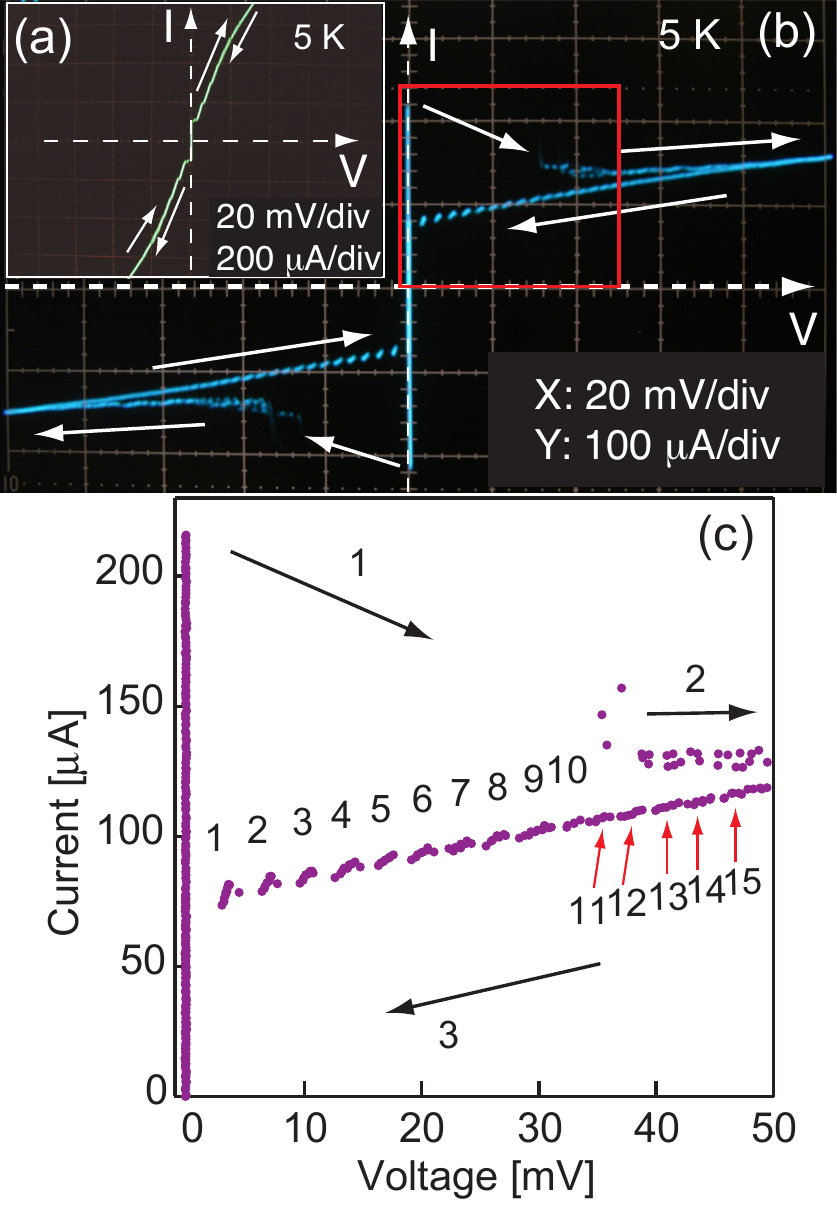}
\end{centering}
\caption{(Color online) $I-V$ curves of the IJJ stack FA\#8. 
(a) An $I-V$ curve after 3 min etching (corresponding to the red dot curve in Fig. \ref{RT.pdf}). 
(b) An $I-V$ curve after 3 more min etching (corresponding to the green solid curve in Fig. \ref{RT.pdf}). 
The white arrows indicate the course of the current ramp. 
(c) Enlarged figure of the box in (b). 
The numbers show indexes of the $N_{\textrm{th}}$ resistive branch. 
}
\label{IV1.pdf} 
\end{figure}

The resistance versus temperature ($R$ - $T$) curves of FA\#8 are shown in Fig. \ref{RT.pdf}. 
The black dashed curve in Fig. \ref{RT.pdf} represents the resistance of the micro bridge before making the two slits, as in Fig. \ref{FIBetching.pdf}(a). 
The red dotted curve in Fig. \ref{RT.pdf} shows the resistance after making the two slits and 3 min of argon ion etching [see Fig. \ref{FIBandmilling.pdf}(c)]. 
One can see that the resistance drastically increases. 
Figure \ref{IV1.pdf}(a) represents the $I-V$ characteristic at this point. 
A tiny voltage jump without hysteresis, which is similar to a weak-link junction, appears. 
This implies that the LSCO IJJ stack starts being formed, but at this point the thickness is almost zero. 
The green solid curve in Fig. \ref{RT.pdf} is after three more min etching (6 min total). 
The resistance increases further, and the $R$ - $T$ curve becomes semiconductive above the superconducting transition temperature. 
Figure \ref{IV1.pdf}(b) and (c) show the $I - V$ curves after etching for 6 min total. 
Resistive multi-branches emerge, and these are quite distinguishable, especially in the low voltage regime of 0 to 30 mV as indexed in Fig. \ref{IV1.pdf}(c). 
These results both the $R - T$ curves and the $I - V$ curves suggest that the depth of the upper slit has increased along the $c$-axis by the argon ion etching and eventually an LSCO IJJ stack is formed as schematically depicted in Fig. \ref{FIBandmilling.pdf}(c).

The superconducting transition temperature $T_{c}$ in Fig. \ref{RT.pdf} is recognized to be 24 K, which is taken from the mid-points of the transitions. 
The Sr concentration of FA\#8 is thus estimated to be $x \approx$ 0.09\cite{Takagi:PRB1989}, and is consistent with the nominal value. 
We note that $T_{c}$ of all three curves does not change, implying that argon ion etching does not affect the carrier density of the crystal. 
This is a great advantage for iterative experiments, such as in Ref. \cite{You:APL2006}, as well as for applications. 

In Fig. \ref{IV1.pdf}(c), we determine that $N_{\textrm{IJJ}}$ is about 15. 
At higer voltage, the gaps between each resistive brach are narrower. 
This is probably attributed to self-heating of each LSCO IJJ, as is the case for BSCCO. 
Assuming $N_{\textrm{IJJ}} \approx 15$, we roughly estimate the etching rate of our argon ion etching machine to be about 3.5 nm/min. 
With proper values of the acceleration voltage and ion beam current, we should be able to etch LSCO crystals with an etch rate of 0.7 nm/min, namely 1 IJJ/min, as already demonstrated for BSCCO by You \textit{et al.}\cite{You:APL2006}. 
Thus this method may let us control $N_{\textrm{IJJ}}$ of an LSCO IJJ stack with an accuracy of single IJJ.

It is interesting that the resistive multi branches are not observed when ramping up the bias current, but observed only while ramping down, as we can see in Fig. \ref{IV1.pdf}(b) and (c). 
After switching [arrow 1 in Fig. \ref{IV1.pdf}(c)], the IJJ stack jumps to around 40 mV and goes up to the maximal voltage ($\sim$ 100 mV) with chaotic behavior [arrow 2 in Fig. \ref{IV1.pdf}(c)]. 
Then it goes down to the superconducting state while showing multi-branches [arrow 3 in Fig. \ref{IV1.pdf}(c)]. 
We have already reported a similar feature of $I-V$ characteristics in a larger ($\approx$ 3 $\mu$m$^{2}$) LSCO IJJ stack\cite{Kubo:PC2008}, and this phenomenon may be attributed to the strong interaction among stacked IJJs\cite{Machida:PRB2004}. 
Further studies are needed to understand this. 


\section{Conclusions}

In conclusion, we successfully fabricated submicron La$_{2-x}$Sr$_{x}$CuO$_{4}$ (LSCO) intrinsic Josephson junction (IJJ) stacks either by focused ion beam (FIB) etching or by an original combined FIB/argon ion beam etching technique. 
To our knowledge, this is the first demonstration of submicron LSCO IJJ stacks. 
About 15 resistive multi-branches appeared in the sample fabricated by the latter technique. 
This method may let us control the number of stacked LSCO IJJs with an accuracy of a single IJJ. 
This method could also be applied to other cuprate materials which are difficult to cleave with Scotch tape, such as rare earth compounds REBa$_{2}$Cu$_{3}$O$_{7-y}$ (RE: rare earth) as well as other Lanthanum compounds.

\begin{acknowledgments}
This work was partially supported by MEXT under Grant No. 1905014. 
The authors thank M. Tachiki and H. Itozaki for the use of the ion etching equipment, K. Takahashi for the help for the machine maintenance, and M. Nagao for discussions. 
Y.K. appreciates \c{C}. Girit for English correction. 
Y.K. was financially supported by NIMS. 
\end{acknowledgments}


\begin{thebibliography}{40}


\bibitem{Kleiner:PRB1994} R. Kleiner and P. M\"{u}ller, Phys. Rev. B \textbf{49} 1327 (1994). 

\bibitem{Inomata:PRL2005} K. Inomata, S. Sato, K. Nakajima, A. Tanaka, Y. Takano, H. B. Wang, M. Nagao, H. Hatano, and S. Kawabata, Phys. Rev. Lett. \textbf{95} 107005 (2005). 

\bibitem{Jin:PRL2006} X.Y. Jin, J. Lisenfeld, Y. Koval, A. Lukashenko, A.V. Ustinov, and P. Muller, Phys. Rev. Lett. \textbf{96} 177003 (2006). 

\bibitem{SXLi:PRL2007} S.-X. Li, W. Qiu, S. Han, Y. F. Wei, X. B. Zhu, C. Z. Gu, S. P. Zhao, and H. B. Wang, Phys. Rev. Lett. \textbf{99} 037002 (2007). 

\bibitem{Kashiwaya:JPSJ2008} H. Kashiwaya, T. Matsumoto, H. Shibata, S. Kashiwaya, H. Eisaki, Y. Yoshida, S. Kawabata, and Y. Tanaka, J. Phys. Soc. Jpn. \textbf{77} 104708 (2008). 

\bibitem{Ota:PRB2009} K. Ota, K. Hamada, R. Takemura, M. Ohmaki, T. Machi, K. Tanabe, M. Suzuki, A. Maeda, and H. Kitano, Phys. Rev. B \textbf{79} 134505 (2009). 

\bibitem{Ueda:JAP2009} S. Ueda, T. Yamaguchi, Y. Kubo, S. Tsuda, Y. Takano, J. Shimoyama, and K. Kishio, J. Appl. Phys. \textbf{106} 074516 (2009). 

\bibitem{Kubo:APEX2010} Y. Kubo, Y. Takahide, S. Ueda, Y. Takano, and Y. Ootuka, Appl. Phys. Exp. \textbf{3} 063104 (2010). 

\bibitem{Kawabata:PRB2004} S. Kawabata, S. Kashiwaya, Y. Asano, and Y. Tanaka, Phys. Rev. B \textbf{75} 014502 (2004). 

\bibitem{Yokoyama:PRB2007} T. Yokoyama, S. Kawabata, T. Kato, and Y. Tanaka, Phys. Rev. B \textbf{76} 134501 (2007). 

\bibitem{Machida:SST2007} M. Machida and T. Koyama, Supercond. Sci. Technol. \textbf{20} S23 (2007). 

\bibitem{Koyama:PC2008} T. Koyama and M. Machida, Physica C \textbf{468} 695 (2008). 

\bibitem{Savelev:PRB2008} S. Savel'ev, A.O. Sboychakov, A. L. Rakhmanov, and F. Nori, Phys. Rev. B \textbf{77} 014509 (2008). 


\bibitem{Kim:SST1999} S.-J. Kim, Y. I. Latyshev, and T. Yamashita, Supercond. Sci. Technol. \textbf{12} 729 (1999). 

\bibitem{Wang:APL2001} H.B. Wang, P.H. Wu, and T. Yamashita, Appl. Phys. Lett. {\bf 78} 4010 (2001). 

\bibitem{You:APL2006}L.X. You, M. Torstensson, A. Yurgens, D. Winkler, C.T. Lin, and B. Liang, Appl. Phys. Lett. {\bf 88} 222501 (2006).

\bibitem{Tamasaku:PRL1992}K. Tamasaku, Y. Nakamura, and S. Uchida, Phys. Rev. Lett. \textbf{69} 1455 (1992). 

\bibitem{Uchida:PRB1996}S. Uchida and K. Tamasaku, Phys. Rev. B \textbf{53} 14558 (1996). 


\bibitem{Martinis:PRB1987}J. M. Martinis, M. H. Devoret, and J. Clarke, Phys. Rev. B \textbf{35} 4682 (1987). 

\bibitem{Wallraff:RSI2003} A. Wallraff, A. Lukashenko, C. Coqui, A. Kemp, T. Duty, and A. V. Ustinov, Rev. Sci. Instrum. \textbf{74}  3740 (2003).


\bibitem{Uematsu:PC2001}Y. Uematsu, \textit{et al.}, Physica C \textbf{362} 290 (2001). 


\bibitem{Kim:IEEE2005}S.-J. Kim, \textit{et al.}, IEEE Trans. Appl. Supercond. \textbf{15} 3782 (2005). 

\bibitem{Kubo:PC2008}Y. Kubo, T. Tanaka, Y. Takahide, S. Ueda, T. Okutsu, A.T.M.N. Islam, I. Tanaka, and Y. Takano: Physica C \textbf{468} 1922 (2008). 

\bibitem{Kitano:PC2009}H. Kitano, K. Ota, K. Ishikawa, M. Itoi, Y. Imai, and A. Maeda, Physica C 470, S838 (2010). 

\bibitem{Mizugaki:JAP2003}Y. Mizugaki, Y. Uematsu, S.-J. Kim, J. Chen, K. Nakajima, T. Yamashita, H. Sato, and M. Naito, J. Appl. Phys. \textbf{94} 2534 (2003). 

\bibitem{Ota:Thesis}K. Ota, Ph. D thesis, The University of Tokyo, 2009. 

\bibitem{FIB}In our system, the etch rate is roughly estimated to be 2 $\mu$m/min for 100 $\mu$m$^{2}$ area of LSCO with the highest emission current mode. 
Needless to say, the etch rate becomes much slower with lower emission current mode which is needed to form a fine 3D structure such as a submicron IJJ stack. 

\bibitem{Tanaka:N1989}I. Tanaka and H. Kojima, Nature (London) \textbf{337} 21 (1989).

\bibitem{Krasnov:PRB2007}V. M. Krasnov, T. Golod, T. Bauch, and P. Delsing, Phys. Rev. B \textbf{76} 224517 (2007).

\bibitem{Takagi:PRB1989}H. Takagi, T. Ido, S. Ishibashi, M. Uota, S. Uchida, and Y. Tokura, Phys. Rev. B \textbf{40} 2254 (1989).

\bibitem{Machida:PRB2004}M. Machida and T. Koyama, Phys. Rev. B \textbf{70} 024523 (2004). 


\end{thebibliography}
\end{document}